\documentclass{iopart}

\usepackage{graphicx}
\usepackage{enumerate}
\usepackage{iopams}

\newcommand{\eg}{\emph{e.g.}}
\newcommand{\se}[1]{\ensuremath{^{\mathrm{#1}}}}
\newcommand{\si}[1]{\ensuremath{_{\mathrm{#1}}}}
\newcommand{\vect}[1]{\ensuremath{\vec{#1}}}
\newcommand{\dagg}{\ensuremath{^{\dagger}}}
\newcommand{\degree}{\ensuremath{^{\circ}}}
\newcommand{\ket}[1]{\ensuremath{\left|{#1}\right>}}

\newcommand{\dirA}{\ensuremath{\tau_1}}
\newcommand{\dirB}{\ensuremath{\tau_2}}
\newcommand{\ketGS}{\ket{\mathrm{0}}}
\newcommand{\ldepth}{\alpha}

\begin{document}
 
\title{Quantum Phases of Trapped Ions in an Optical Lattice}
\author{R Schmied, T Roscilde, V Murg, D Porras and J I Cirac}
\address{Max-Planck-Institut f\"ur Quantenoptik, 85748 Garching, Germany}
\ead{Roman.Schmied@mpq.mpg.de}

\begin{abstract}
We propose loading trapped ions into microtraps formed by an optical lattice. For harmonic microtraps, the Coulomb coupling of the spatial motions of neighboring ions can be used to construct a broad class of effective short-range Hamiltonians acting on an internal degree of freedom of the ions. For large anharmonicities, on the other hand, the spatial motion of the ions itself represents a spin-1/2 model with frustrated dipolar XY interactions. We illustrate the latter setup with three systems: the linear chain, the zig-zag ladder, and the triangular lattice. In the frustrated zig-zag ladder with dipolar interactions we find chiral ordering beyond what was predicted previously for a next-nearest-neighbor model. In the frustrated anisotropic triangular lattice with nearest-neighbor interactions we find that the transition from the one-dimensional gapless spin-liquid phase to the two-dimensional spiraling ordered phase passes through a gapped spin-liquid phase, similar to what has been predicted for the same model with Heisenberg interactions. Further, a second gapped spin-liquid phase marks the transition to the two-dimensional N\'eel-ordered phase.
\end{abstract}

\submitto{\NJP}
 

\section{Introduction}

The merging of atomic physics and condensed matter physics has opened exciting new perspectives for the creation and manipulation of quantum states of matter. For example, cold atoms in optical lattices are a setup which allows experimentalists to simulate quantum many-body lattice systems \cite{Bloch2007,Lewenstein2007}. In recent years, trapped ions have also been proposed as an experimental system where a rich variety of quantum many-body models can be implemented \cite{Mintert2001,Wunderlich2001,Porras2004,Porras2004b,Deng2007}. The advantage of trapped ions is that we profit from the technology that has been developed for quantum information processing \cite{Kielpinski2002,Leibfried2003,Gulde2003,SchmidtKaler2003,Stick2006,Pearson2006}. In particular, it is possible to measure and manipulate the system at the single-particle level, due to the large inter-particle distances and the optical accessibility of many electromagnetic traps.

So far, previous proposals fall into two categories: (i) effective-spin models, in which the motion of the ions induces an effective interaction between internal levels \cite{Wunderlich2001,Porras2004,Deng2005}, and (ii) interacting-Boson models, in which the vibrations (phonons) play the role of interacting particles \cite{Porras2004b,Deng2007}. In this work we show that these proposals may be scaled up to a wide range of spatial dimensions, geometries, and particle interactions by placing the ions in a regular structure. This structure may be induced by optical lattices or by an array of ion microtraps.

The presence of the trapping structure offers us a way to increase the dimensionality of the system beyond the traditional linear Paul trap scheme. Further, it also opens up new possibilities for controlling particle interactions. On one hand, since ions are tightly confined by individual trapping potentials, their motion in any spatial direction can be used to induce spin--spin interactions which decay with the third power of the distance. This situation is to be compared with the case of Coulomb crystals confined in an overall trapping potential, where vibrational phonons with small wavevector, corresponding to highly collective vibrations, have a low-energy spectrum and induce infinite-range spin--spin interactions only \cite{Porras2004}. On the other hand, in the optical lattice trapping scheme, we can control the anharmonicities of the trapping potentials and engineer effective phonon--phonon interactions. Thus, vibrational degrees of freedom may follow a Bose--Hubbard model along the lines of \cite{Porras2004b}.

The realization of a periodic trapping potential with an optical lattice is a standard experimental method.
In this work we focus on the study of the quantum many-body phases which arise in such experimental setups. We discuss two limiting cases. First, the case of small anharmonicities, and the effective quantum spin models which can be implemented by using the ion motion as a medium for inducing effective spin interactions among internal degrees of freedom. Second, we discuss in more detail the ground vibrational state in trapping schemes in which large anharmonicities induce phonon--phonon interactions and allow mapping hard-core phonons onto $S=\frac12$ spins. We focus on the case of periodic trapping potentials with interesting geometries, e.g., the triangular lattice. This case provides us with a playground for studying frustrated quantum magnets in a clean experimental setup. In this way, trapped ions allow us to implement models which have attracted much attention in condensed matter physics.

\section{Microtraps}
\label{sec:microtraps}

The starting point for all of the models in this work is a set of ions trapped in individual microtraps. Such traps have been proposed in the literature to be micro-fabricated, e.g., on surfaces \cite{Chiaverini2007}, which allows for arbitrary geometries of the resulting ion ``crystal'' and, in principle, gives the experimentalist great control over the effective anharmonicity of the spatial ion motion near the microtrap minima.

We propose an alternative scheme for creating microtrap potentials, postponing detailed calculations to a further publication \cite{SchmiedLeibfried}. Optical lattices have been used extensively in experiments with cold atoms, which interact mostly on short length scales and can populate $d$-dimensional lattices densely. If a $d$-dimensional Wigner crystal, assembled in an electromagnetic trap, is subjected to a sufficiently strong optical lattice potential, the spatial motion of each ion is restricted to a region near the minimum of the lattice well closest to the crystal equilibrium position. However, the effective amplitude of a realistic optical lattice is much smaller than the repulsion of two ions in the same or even neighboring lattice wells. Therefore, as opposed to experiments with cold atoms, the optical lattice will be very sparsely populated with ions, but their strong interactions over large distances are nevertheless sufficient for experimentally exploring the models presented in \sref{sec:models}. Depending on the stiffness of the Wigner crystal used as a starting point, the optical lattice does not need to be so strong as to trap the ions by itself; it is sufficient that the zero-point motion of the Wigner crystal is reduced by the optical lattice (since the optical lattice increases the effective mass of the ions) to the point where each ion is constrained to moving within a single lattice minimum. In a one-dimensional optical lattice potential $V(z)=\ldepth E\si{R}\sin^2(q z)$, with the recoil energy $E\si{R}=\hbar^2q^2/(2m)$, the harmonic frequency of the ions moving in their local microtraps is $\hbar\omega\si{mt}=2\sqrt{\ldepth}E\si{R}$. Recoil energies are typically tens to hundreds of kHz; with an optical lattice depth of hundreds of recoil energies ($\ldepth\gtrsim100$) it is thus feasible to reach microtrap frequencies $\omega\si{mt}$ far in excess of the ion trap frequency $\Omega$, the latter being typically on the same order as $E\si{R}$ but possibly as low as necessary. Lowering the ion trap frequency significantly below the recoil energy of the optical lattice is an alternative to large values of $\ldepth$ for achieving well-defined microtraps. The energy scale of the couplings between the vibrational motion of different ions, separated by a typical distance $d_0$, is $E\si{t}=E\si{C}(q d_0)^{-2}\ldepth^{-1/2}$, with the Coulomb energy scale $E\si{C}=e^2/(4\pi\varepsilon_0 d_0)$ [see also \eref{eq:tunnel_gen}]. The low filling fraction of the optical lattice ($q d_0\gg1$) reduces the very large Coulomb energy scale, typically tens or hundreds of GHz, to below the microtrap energy scale $\hbar\omega\si{mt}$: in a harmonic trap, the typical distance between ions is of order $d_0\approx [e^2/(2\pi\varepsilon_0 m\Omega^2)]^{1/3}$, which gives $E\si{t}\approx \hbar\omega\si{mt}(\hbar\Omega/E\si{R})^2(8\ldepth)^{-1} \ll\hbar\omega\si{mt}$. Keeping this coupling energy scale $E\si{t}$ below the microtrap energy scale is necessary in order to prevent processes that change the number of vibrational excitations, which are assumed to be absent throughout the rest of this work.

As we have already mentioned, many interesting models require significant phonon--phonon interactions (see \sref{sec:spinmodels}). The anharmonicity of the Coulomb interaction in a Wigner crystal is insufficient for observing these phenomena, and a harmonic normal-mode analysis captures the full physics of low-energy ion motion. The reason for this is that the length scale $\zeta$ of ion vibrations is only a fraction of the optical-lattice wavelength and thus much smaller than the typical distance $d_0$ between neighboring ions; quartic terms of the Coulomb interaction, which give rise to phonon--phonon interactions, are smaller than the harmonic terms (energy scale $E\si{t}$) by a factor $(\zeta/d_0)^2\ll1$ \cite{Roos2007b}. However, we can make use of the natural anharmonicity of the sinusoidal optical lattice potential to introduce significant phonon--phonon interactions. This anharmonicity is close to the recoil energy $E\si{R}$ for $\ldepth\gtrsim10$. The ratio of the anharmonicity to the coupling energy scale is $E\si{R}/E\si{t} = \hbar\omega\si{mt}/E\si{t}/(2\sqrt{\ldepth})$; with the above assumption of $E\si{t}\ll\hbar\omega\si{mt}$, this ratio is typically larger than one, giving the effective hard-core repulsion between phonons which is the basis of the models presented in \sref{sec:models}. But it must be noted that the \emph{sign} of the anharmonicity is negative for a strong monochromatic optical lattice. If the lattice is sufficiently strong, the ions are forced to the lattice minima, where the quartic anharmonicity is negative and leads to an effective phonon--phonon \emph{attraction}; effective phonon models with attractive interactions have very simple ground states and do not merit experimental study on a quantum simulator. Much more interesting phases result from \emph{repulsive} phonon--phonon interactions, which can be achieved by using an optical superlattice \cite{Foelling2007}. Briefly, if a second optical lattice with half the wavelength of the first one is superposed with the same phase and a reduced amplitude $\ldepth/64<\ldepth'<\ldepth/16$, then the anharmonicity at the superlattice minima can be made positive while still retaining the same spatial arrangement of microtrap minima; further, the amplitude ratio $\ldepth'/\ldepth$ is a sensitive parameter for adjusting the ratio between the energy scales of phonon tunneling ($E\si{t}$) and repulsion ($E\si{R}$). In what follows we assume repulsive phonon--phonon interactions, irrespective of how the microtraps are put in place.

\section{Varying the Anharmonicity}

This section describes three regimes of anharmonicities of the microtraps. If the anharmonicities are negligible, the internal degrees of freedom of trapped ions can be made to interact via dipolar phonon-mediated couplings. At the opposite end, for strong anharmonicities the quantized vibrational modes themselves describe a $S=\frac12$ spin model with XY interactions. In the intermediate regime, the vibrational motion of the ions can be effectively described by an interacting Bose--Einstein condensate.

\subsection{No anharmonicity: short-range lattice Hamiltonians}
\label{sec:internalDoF}

In \cite{Porras2004} a model is elaborated where internal states of trapped ions represent effective $S=\frac12$ spin degrees of freedom. In this model, the vibrational modes of a self-assembled ion crystal in a harmonic trap mediate variable interactions between the effective spins, allowing for controllable interactions in all spin directions. A drawback of the proposed architecture is that the longitudinal vibrational modes, where the ion crystal moves as a whole in the ion trap, are naturally slow (``floppy'' modes) and lead to long-range ferromagnetic interactions. This limits the generality of the Hamiltonians that can be constructed with such crystals of trapped ions.

If, however, local microtraps are added to the confining potential, then this problem can be overcome by increasing the potential energy curvature in \emph{all} directions of motion of the individual trapped ions. In the simplest case, adding the same isotropic harmonic microtrap potential to each ion, the dispersion and shape of normal modes of the ion crystal are unchanged except for an overall shift in frequency corresponding to the curvature of the microtraps. Such a frequency shift renders the vibrational modes ``stiff'' for all three spatial directions of each ion. Consequently the effective spin-$\frac12$ degrees of freedom of the trapped ions can be made to interact through Ising, XY, or XYZ couplings, all of which have adjustable amplitudes and decay as $r^{-3}$ since they stem from the dipole--dipole interactions of the ionic vibrations. \cite{Porras2004} further shows that if the phonon modes have only finite stiffness, then next-nearest-neighbor interactions can be suppressed, approaching effective nearest-neighbor Hamiltonians on arbitrary lattice geometries.

\subsection{Small anharmonicity: interacting Bose--Einstein condensate}
\label{sec:BEC}

As discussed in \sref{sec:microtraps}, the vibrational motion of a self-assembled ion crystal in a global trap is very well approximated by a harmonic model, as was used in Sec.~\ref{sec:internalDoF}. At low energies, cubic terms of the Coulomb interaction are unimportant because the three-phonon processes they describe are highly nonresonant; quartic terms, on the other hand, can be neglected because of their small amplitudes.

This picture changes when microtraps are added to the local potentials of the individual ions. In addition to the harmonic confinement, these microtraps can add significant anharmonicities $U_{\alpha}$ to the ionic vibrations. As discussed in \cite{Porras2004b,Deng2007}, in the absence of fast phonon decay channels the spatial motion of trapped ions thus behaves like a Bose--Hubbard model of phonons, with Hamiltonian
\begin{equation}
	\label{eq:BHham}
	\mathcal{H}\si{BH} = \sum_{\langle \alpha,\beta\rangle} t_{\alpha,\beta}(a_{\alpha}\dagg a_{\beta} + \mathrm{h.c.})
	+ \sum_{\alpha} U_{\alpha} n_{\alpha}(n_{\alpha}-1)
	+ \sum_{\alpha} V_{\alpha} n_{\alpha},
\end{equation}
where the indices $\alpha,\beta$ run over the $3N$ eigen-directions of motion of the $N$ ions. Here we assume that the quartic phonon--phonon interaction terms (anharmonicities) do not couple different modes, even if the modes are localized on the same ion. This decoupling occurs naturally in a cubic optical lattice; but, irrespective of the microtrap geometry, we will assume throughout this work that of the three directions of motion of each ion, only one is ``active'' and contributing to the models we wish to study, while the other two modes are off-resonant and do not influence the low-energy dynamics of the active mode. This is achieved with strongly anisotropic prolate microtraps, where one slow direction of motion is decoupled from the two fast directions of motion. (Extensions of the present models to two or three ionic degrees of freedom are straightforward and do not introduce new physics if the degrees of freedom are separable.) For example, in a three-dimensional cubic optical lattice where one standing wave is much weaker than the other two, every lattice minimum constitutes such a prolate microtrap.

Calling $\vect{m}_{\alpha}$ the direction of motion of the $\alpha\se{th}$ ion, the effective tunneling matrix elements of the phonons are of the dipolar form
\begin{equation}
	\label{eq:tunnel_gen}
	t_{\alpha,\beta} = \frac{\zeta_{\alpha} \zeta_{\beta} e^2}{8\pi\varepsilon_0 r_{\alpha,\beta}^3} [\vect{m}_{\alpha}\cdot\vect{m}_{\beta}
	-3(\vect{n}_{\alpha,\beta}\cdot\vect{m}_{\alpha})(\vect{n}_{\alpha,\beta}\cdot\vect{m}_{\beta})],
\end{equation}
where $e^2/(4\pi\varepsilon_0)$ is the electrostatic coupling strength, $\vect{r}_{\alpha,\beta}=\vect{r}_{\beta}^{(0)}-\vect{r}_{\alpha}^{(0)}$ are the equilibrium inter-particle spacings, and $\vect{n}_{\alpha,\beta}=\vect{r}_{\alpha,\beta}/r_{\alpha,\beta}$. The length scales $\zeta_{\alpha}$ are those of the harmonic-oscillator motion in the quadratic component of the local microtrap potentials, related to the local potentials as $\zeta_{\alpha} = \hbar/\sqrt{m V_{\alpha}}$, where $m$ is the mass of the ions. In a homogeneous lattice of microtraps, where all $\vect{m}_{\alpha}$ are equal, the angular dependence of~\eref{eq:tunnel_gen} is proportional to the Legendre polynomial $t_{\alpha,\beta}\propto-P_2(\vect{n}_{\alpha,\beta}\cdot\vect{m})$. In the following sections, we will focus our attention on the central region of the trap, where the tunneling coefficients $t_{\alpha,\beta}$ and local potentials $V_{\alpha}$ are sufficiently isotropic to be approximated by a translationally invariant model.

In the absence of anharmonicities ($U_{\alpha}=0$ $\forall\alpha$), the Hamiltonian~\eref{eq:BHham} can be easily diagonalized. Irrespective of the signs of the tunneling matrix elements $t_{\alpha,\beta}$, there will be a lowest-energy normal mode $z$ (possibly degenerate by symmetry). The ground state in the canonical ensemble of fixed phonon number $n$ is simply $(n!)^{-1/2}(b_z\dagg)^n\ketGS$, which can be interpreted as a Bose--Einstein condensate (BEC) of phonons in mode $z$. 

The BEC analogy can be extended to small but nonzero phonon--phonon interactions $U_{\alpha}$. With the canonical transformation $a_{\alpha} = \sum_k \Gamma_{\alpha,k}b_k$ that diagonalizes the harmonic part of the Hamiltonian (i.e., describes the normal modes), the Hamiltonian is re-expressed as 
\begin{equation}
	\label{eq:BHhamNM}
	\mathcal{H}\si{BH} = \sum_k \tilde{V}_k b_k\dagg b_k + \sum_{k_1,k_2,k_3,k_4} U_{k_1,k_2}^{k_3,k_4} b_{k_1}\dagg b_{k_2}\dagg b_{k_3} b_{k_4},
\end{equation}
with the coefficients $U_{k_1,k_2}^{k_3,k_4}=\sum_{\alpha} U_{\alpha} \Gamma_{\alpha,k_1}^*\Gamma_{\alpha,k_2}^*\Gamma_{\alpha,k_3}\Gamma_{\alpha,k_4}$ describing interactions between normal modes which deplete the BEC.
When approximating the central region of the ion crystal by a quasi-infinite periodic $d$-dimensional lattice of $N$ ions, the normal modes become Fourier modes $b_{\vect{k}} = N^{-1/2}\sum_j a_j \exp(-i\vect{k}\cdot\vect{r}_j)$.
The interaction of these modes conserves linear momentum:
\begin{equation}
	\mathcal{H} = \sum_{\vect{k}} \hat{t}(\vect{k}) n_{\vect{k}}
	+ U N^{-1} \sum_{\vect{k}_1,\vect{k}_2,\vect{k}_3,\vect{k}_4}
	b_{\vect{k}_1}\dagg b_{\vect{k}_2}\dagg b_{\vect{k}_3} b_{\vect{k}_4}
	\delta_{(\vect{k}_1+\vect{k}_2),(\vect{k}_3+\vect{k}_4)},
\end{equation}
where $\hat{t}(\vect{k}) = \sum_{\vect{\delta}} t(\vect{\delta}) e^{i\vect{k}\cdot\vect{\delta}}$ are the Fourier transformed tunneling couplings, and $n_{\vect{k}} = b_{\vect{k}}\dagg b_{\vect{k}}$. In this approximation, a small interaction $U$ partly depletes the BEC at $\vect{k}_z$ and broadens the structure factor around this momentum component. The condensate depletion fraction is then \cite{PethickSmith}
\begin{equation}
	\label{eq:condensatedepletion}
	\frac{n\si{ex}}{n_0} = \frac{1}{N} \sum_{\vect{k}\neq\vect{k}_z}
	\frac12\left( \frac{\hat{t}(\vect{k})+2U n_0}
	{\sqrt{[\hat{t}(\vect{k})]^2 + 4U n_0 \hat{t}(\vect{k})}}-1 \right),
\end{equation}
where $n_0=n/N$ is the undepleted BEC density, and the dispersion relation is shifted such that $\hat{t}(\vect{k}_z)=0$.

\subsection{Large anharmonicity: spin-$\frac12$ XY models}
\label{sec:spinmodels}

In the limit of strong repulsion $|U_{\alpha}| \gg |\sum_{\beta} t_{\alpha,\beta}|$ $\forall \alpha$ and for low phonon filling $n<1$, double phononic occupancy is strongly suppressed, and the system of phonons attains the hardcore limit, where phonons can be conveniently mapped onto $S=\frac12$ spins via the Holstein--Primakoff transformation $a_{\alpha}\dagg\to S_{\alpha}^+$, $a_{\alpha}\to S_{\alpha}^-$, $n_{\alpha} \to S_{\alpha}^z+\frac12$. For each ion, the vibrational ground state $\ket{0}$ is thus mapped onto the spin state $\ket{\downarrow}$, and the lowest excited state $\ket{1}$ onto the state $\ket{\uparrow}$. The resulting spin Hamiltonian has the form of an XY model with long-range dipolar couplings and a site-dependent field:
\begin{equation}
	\label{eq:spinham}
	\mathcal{H}\si{S} = 2\sum_{\langle \alpha,\beta\rangle} t_{\alpha,\beta}(S_{\alpha}^x S_{\beta}^x + S_{\alpha}^y S_{\beta}^y)
	+ \sum_{\alpha} V_{\alpha} S_{\alpha}^z.
\end{equation}
Any spin--spin couplings for the $z$ components must derive from off-site density--density interactions between the phonons, namely from terms of the form $n_in_j$. The only source of such interactions is the fourth-order terms of the Coulomb interaction between the ions; as discussed in \sref{sec:microtraps}, they are smaller than the XY interaction by a factor of $(\zeta/d_0)^2$ and therefore negligible. Nonetheless the implementation of the Hamiltonian of~\eref{eq:spinham} is of significant interest, because it offers the possibility of exploring the rich physics of frustrated XY spin models in various dimensions, with a fully tunable frustration. Detailed discussions of several such spin models and associated quantum phases and phase transitions are provided in the next section.

\section{Frustrated XY Spin-$\frac12$ Models}
\label{sec:models}

In this section we discuss three frustrated XY lattice spin models belonging to the family described by the Hamiltonian of~\eref{eq:spinham} and realizable in linearly or planarly trapped ions. We henceforth assume for simplicity uniform local potentials $V_{\alpha}=V$ in the trapping region; under the condition of preparing the system with a well-defined number of phonons, we can then discard the field term in~\eref{eq:spinham}. Moreover we focus on a half-filled system of hardcore bosons, corresponding to the zero-magnetization sector of the Hilbert space for the spin system. In the case of bipartite lattices, the ground state of an XY antiferromagnet in this zero-magnetization sector can be rigorously shown to coincide with that of a (more common) XY antiferromagnet without magnetization constraints and in zero field \cite{Mattis1979}. As for frustrated lattices, this remains certainly true for finite-size systems, where no spontaneous breaking of the $Z_2$ symmetry of the Hamiltonian can occur. 

We have focused on spin models with XY interactions, featuring full frustration for two equivalent (and non-commuting) spin components, a paradigm of quantum frustration. When using internal states of the ions to encode the spin variable (\sref{sec:internalDoF}), the effective spin--spin interaction which arises is typically of Ising-like symmetry, and full frustration appears only for one spin component at a time. Fine tuning of the system parameters would be required to obtain a rotationally invariant Hamiltonian, \eg\ in the XY plane. On the other hand, when spins are encoded in the phononic states of each ion (\sref{sec:spinmodels}), the XY rotational symmetry is a robust property of the Hamiltionan, as it stems from the conservation of the number of phonons.

\paragraph{Linear chain.}

The Hamiltonian of~\eref{eq:spinham} on a linear chain, obtained by strong transverse confinement of the ions, realizes an $S=\frac12$ one-dimensional XY antiferromagnet with dipolar interactions. If we neglect all interactions except nearest-neighbor ones, we recover an exactly solvable model \cite{Lieb1961} with power-law decaying correlations in the ground state, $\langle S_i^z S_{i+r}^z \rangle \sim r^{-2}$ and $\langle S_i^x S_{i+r}^x \rangle \sim r^{-1/2}$. Long-range dipolar tunneling simply modifies the decay exponents \cite{Deng2005}, leading to a faster decay of $\langle S_i^x S_{i+r}^x \rangle$ and to a slower decay of $\langle S_i^z S_{i+r}^z \rangle$.

\paragraph{Zig-zag ladder.}

\begin{figure}
	\centering
	\includegraphics[width=8cm]{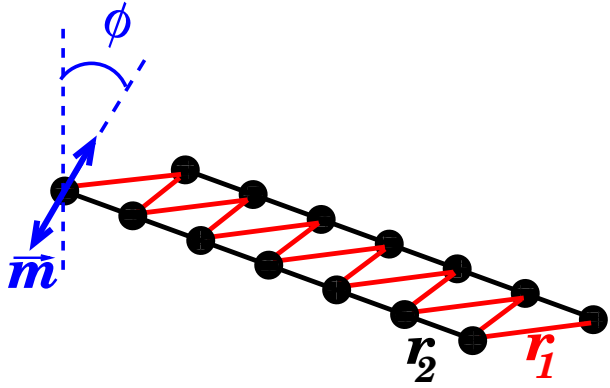}
	\caption{The zig-zag ladder with inter-chain distances $r_1=d_0\sqrt{1+\xi^2}$ and intra-chain distances $r_2=2d_0$ . The motions of the ions in the direction $\vect{m}$ of the blue arrow are coupled to form a spin lattice model.}
	\label{fig:zigzag_schematic}
\end{figure}

A richer physics emerges in the case of a zig-zag ladder of ions, which develops naturally for weaker transverse confinement of the ions \cite{Birkl1992,Schiffer1993}. Assuming longitudinal inter-ion spacing $d_0$ and a zig-zag amplitude $\xi\times d_0$, inter-chain neighboring ions are distant by $r_1=d_0\sqrt{1+\xi^2}$ and intra-chain neighbors by $r_2=2d_0$ (see \fref{fig:zigzag_schematic}). The relevant phonon direction of motion $\vect{m}$ in~\eref{eq:tunnel_gen} is taken perpendicular to the main trap axis and at an angle $\phi$ from the ion plane normal.
Hence the ratio between the dominant intra-chain ($t_2$) and inter-chain ($t_1$) couplings is
\begin{equation}
	\label{eq:Rfromxi}
	R = \frac{t_2}{t_1} = \frac18 \times \frac{(1+\xi^2)^{5/2}}{1-\frac12\xi^2(1-3\cos2\phi)}.
\end{equation}
In trapped-ion experiments with cylindrically symmetric traps, at $\xi \approx 0.965$ spontaneously generated zig-zag ladders deform into helices \cite{Schiffer1993}. For $\phi=0$ (out-of-plane vibrational motion) this limits the range of easily accessible coupling ratios to $R \lesssim 0.335$. In practice, larger amplitudes $\xi$ (and hence a larger $R$) can be engineered by breaking the cylindrical symmetry of the trap and constraining the ions to two dimension, or by microtrap stabilization \cite{Cirac2000}. However, choosing $\phi=\frac{\pi}{2}$ instead (in-plane vibrational motion) gives access to all ratios $R>1/8$ for $\xi<1/\sqrt{2}$. In what follows we therefore assume in-plane vibrational motion. We use $\xi$ and $R$ interchangeably through~\eref{eq:Rfromxi}.

\begin{figure}
	\centering
	\includegraphics[width=12cm]{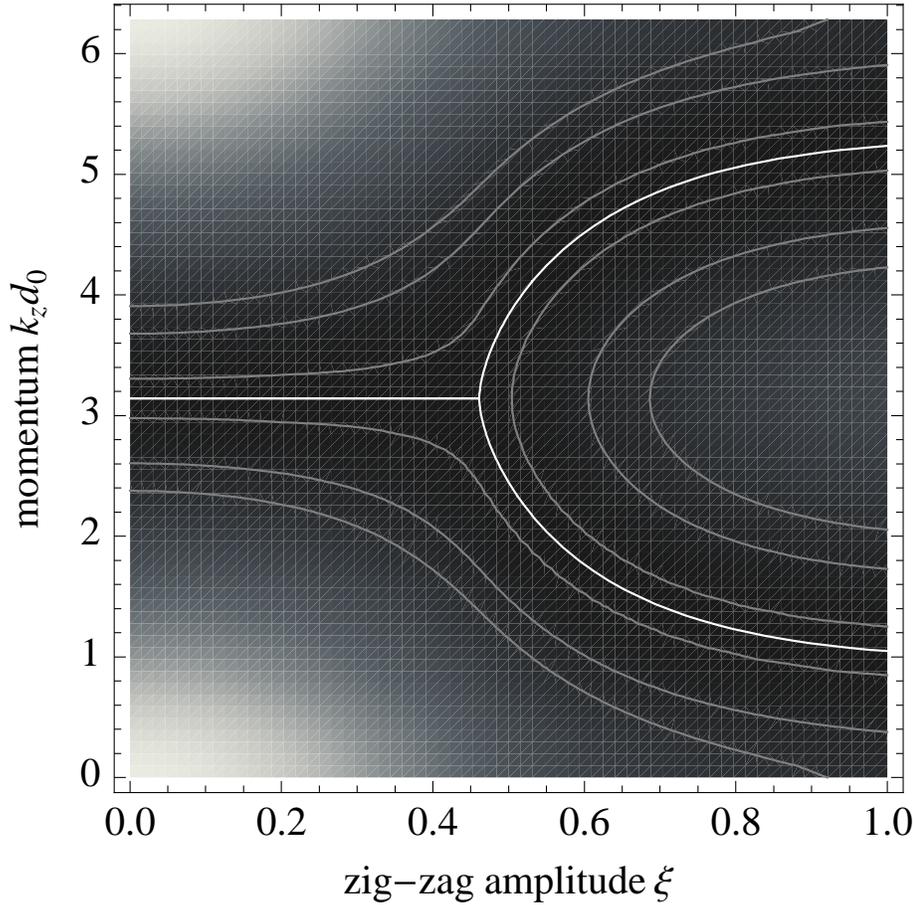}
	\caption{Dispersion $\hat{t}(k)$ of normal modes in an infinite zig-zag ladder with dipolar couplings ($\phi=\frac{\pi}{2}$). The white line traces the minimum as a function of the zig-zag amplitude $\xi$. Lighter shades indicate higher harmonic frequencies; three contours of equal $\hat{t}(k)$ are plotted in grey.}
	\label{fig:zigzag_dispersion}
\end{figure}

\Fref{fig:zigzag_dispersion} shows the normal-mode dispersion relation in the zig-zag ladder. For small zig-zag amplitudes $\xi<\hat{\xi}\si{dipole}\se{classical}=0.461$ ($R<0.353$) the normal-mode dispersion $\hat{t}(k)$ has a unique minimum at $k_z d_0=\pi$, corresponding to N\'eel order; but for $\xi>\hat{\xi}\si{dipole}\se{classical}$ this minimum bifurcates into two symmetric minima at $k_z^{\pm} d_0=\pi\pm q(\xi)$, corresponding to spiral order. This is to be compared with a similar result for the system with intra-chain and inter-chain nearest-neighbor couplings only ($\{t_1,t_2\}$ system), where the transition from N\'eel to spiral order occurs at a smaller value of $R=1/4$ in the limit of classical XY spins \cite{Harada1984}. A spiraling ground state is endowed both with spontaneous magnetic order, namely the breaking of the rotational symmetry in spin space, and with \emph{chiral} order, corresponding to the choice of helicity of the spiraling state.

\begin{figure}
	\centering
	\includegraphics[width=12cm]{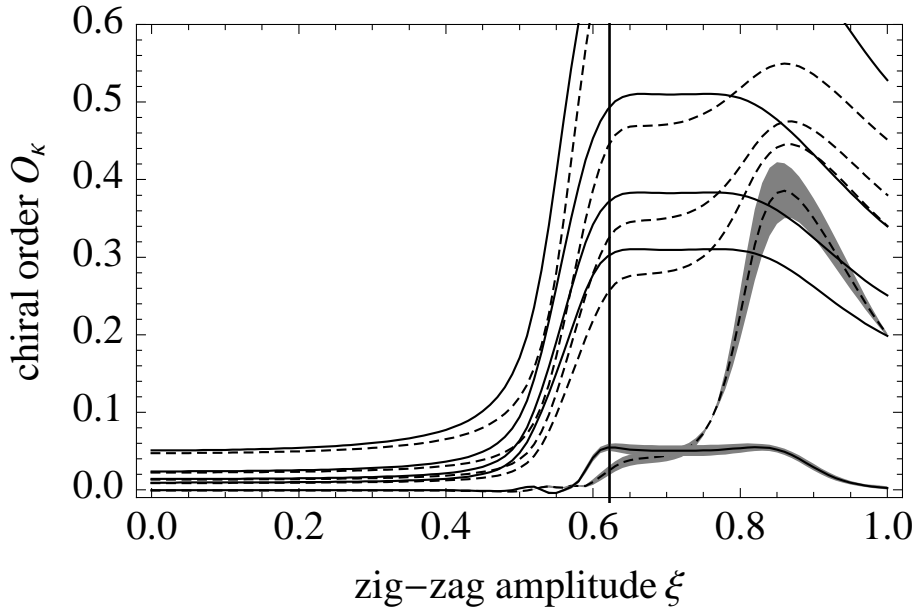}
	\caption{The chiral order parameter $O_{\kappa} $ from~\eref{eq:chiralorder_b} in the ground state of the zig-zag ladder ($\phi=\frac{\pi}{2}$), for the $\{t_1,t_2\}$ model (solid lines) and long-range dipolar interaction (dashed lines), computed from exact diagonalizations (Arnoldi) with $L=8, 12, 16, 20$ spins (top to bottom) and open boundary conditions. The grey bands indicate $2\sigma$ regions of the extrapolation to infinite system size by least-squares fits including $L^{-1}$ and $L^{-2}$ corrections. The vertical line indicates the transition point to long-range chiral order in the $\{t_1,t_2\}$ model, estimated from DMRG calculations \cite{Hikihara2001}.}
	\label{fig:zigzag_chiral}
\end{figure}

In a quasi-1D system we define the chirality as \cite{Hikihara2001}
\begin{equation}
	\label{eq:kappa}
	\kappa_i = 4(S_i^x S_{i+1}^y - S_i^y S_{i+1}^x),
\end{equation}
where the sites $i$ and $i+1$ belong to different chains. A striking result in quantum systems is that chiral order can survive even when magnetic order disappears. In fact, the $\{t_1,t_2\}$ XY zig-zag ladder with $S=\frac12$ is found in \cite{Hikihara2001} to exhibit power-law decaying spin--spin correlations for all values of $R$, but it is also found to develop long-range chiral order for $R>1.26$. We extend the calculation of \cite{Hikihara2001} to an XY ladder with dipolar interactions, making use of exact diagonalization up to sizes $L=20$. We define an order parameter for the chiral phase as the averaged chiral correlation over all pairs of the system:
\numparts
\begin{eqnarray}
	\label{eq:chiralorder_a}
	O_{\kappa}^{\Delta} & = \frac{1}{L-1-|\Delta|} \sum_i \langle \kappa_i \kappa_{i+\Delta} \rangle\\
	\label{eq:chiralorder_b}
	O_{\kappa} & = \frac{1}{2L-3} \sum_{\Delta=-(L-2)}^{L-2} O_{\kappa}^{\Delta}.
\end{eqnarray}
\endnumparts
\Fref{fig:zigzag_chiral} shows this chirality order parameter as a function of the zig-zag amplitude. For the $\{t_1,t_2\}$ model, we confirm a relatively sharp transition to a chirally ordered state for $\xi\approx0.59$ (corresponding to $R\approx0.9$), consistent with the transition of \cite{Hikihara2001}. The full long-range dipolar model essentially retains this transition, shifted to slightly higher zig-zag amplitudes ($\xi\approx0.63$), along with what seems like a reorientation transition around $\xi\approx0.8$ with significantly higher chiral order, which is not observed in the classical limit. At present we do not provide a description of the phase $0.8<\xi<1$, but we point out that despite the perturbative appearance of the dipolar interaction tails, they change the phase diagram of this zig-zag ladder drastically. This effect is currently the subject of further study.

\paragraph{Triangular lattice.}

\begin{figure}
	\centering
	\includegraphics[width=12cm]{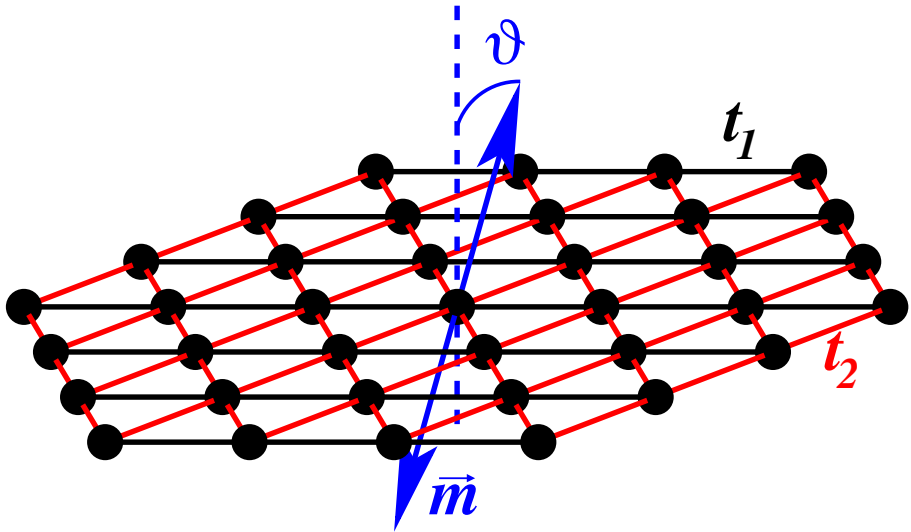}
	\caption{Ion vibrational motion in the triangular lattice. The black (red) lines show the $t_1$ ($t_2$) interactions in the \dirA\ (\dirB) directions. Blue is a possible direction of motion $\vect{m}$, perpendicular to the \dirA\ direction and forming an angle $\vartheta$ with the surface normal (dashed line).}
	\label{fig:triangularmove}
\end{figure}

When put in a planar trap, ions self-assemble into a triangular-lattice Wigner crystal \cite{Mitchell1998}. Adding an optical lattice to this setup, the degrees of freedom associated with the strongly non-linear vibrational motion in the lattice minima can realize a large family of spatially anisotropic XY triangular antiferromagnets. Unlike the case of the zig-zag ladder, we consider fixed average ion positions here, and the model parameters are varied by tilting the direction of motion of the ions $\vect{m}$ with respect to the lattice normal. We further assume that $\vect{m}$ remains perpendicular to one of the lattice directions, indicated by \dirA, and makes an angle $\vartheta$ with the lattice normal (see \fref{fig:triangularmove}).

\begin{figure}
	\centering
	\includegraphics[width=12cm]{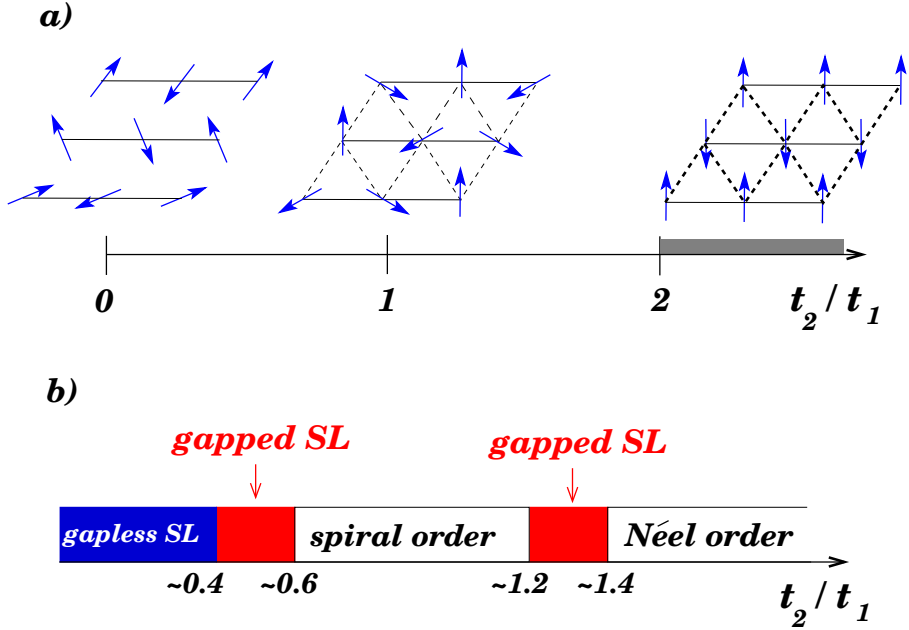}
	\caption{Phase diagram of the $S=\frac12$ XY antiferromagnet on the spatially anisotropic triangular lattice ($\{t_1,t_2\}$ model). a) Classical spin ordering: decoupled N\'eel-ordered horizontal (\dirA) chains for $t_2/t_1=0$; three-sublattice ordering for $t_2/t_1=1$; N\'eel order on the $t_2$ lattice for $t_2/t_1\ge2$. b) Proposed quantum phase diagram for $S=\frac12$ spins, including three spin-liquid (SL) phases.}
	\label{fig:PhD}
\end{figure} 
  
For $\vartheta=0$ the model of~\eref{eq:spinham} realizes an isotropic triangular XY antiferromagnet with dipolar couplings, and it is maximally frustrated. For this system the minima of the normal-mode dispersion are on the corners of the hexagonal first Brillouin zone, corresponding in the classical limit to the well-known three-sublattice ordered state with 120\degree\ angles between the spin directions on the sublattices [see \fref{fig:PhD}(a)].

A nonzero angle $\vartheta$ produces a spatial anisotropy in the couplings, leaving the intra-chain couplings along the \dirA-axis unchanged while modifying the transverse (inter-chain) couplings. The value $\vartheta\approx31\degree$ produces the minimum effective interaction between the \dirA-chains, in the sense that the normal-mode dispersion $\hat{t}(\vect{k})$ is least corrugated in the direction perpendicular to \dirA. We note that $\vartheta=31\degree$ is also the point where the BEC depletion of~\eref{eq:condensatedepletion} grows fastest with $U$, approximating the 1D limit where the BEC is fully depleted for any nonzero interaction $U$. At $\vartheta\approx42\degree$ the nearest-neighbor inter-chain coupling vanishes, while residual weaker couplings to further neighbors survive due to the long-range nature of the dipolar interaction. On the opposite end, a value of $\vartheta=\frac{\pi}{2}$ maximizes the inter-chain nearest-neighbor interaction. Thus by rotating the direction of vibration $\vect{m}$ with respect to the lattice plane we produce a similar effect to that produced in the zig-zag ladder by deformation of the Coulomb crystal. This different method is necessary because tuning the interaction coefficients in the effective spin Hamiltonian by mechanical deformation of the ion crystal is problematic, due to the significant stiffness of a two-dimensional Wigner crystal. The two approaches could possibly be combined in order to cover a larger range of system parameters.

\begin{figure}
	\centering
	\includegraphics[width=12cm]{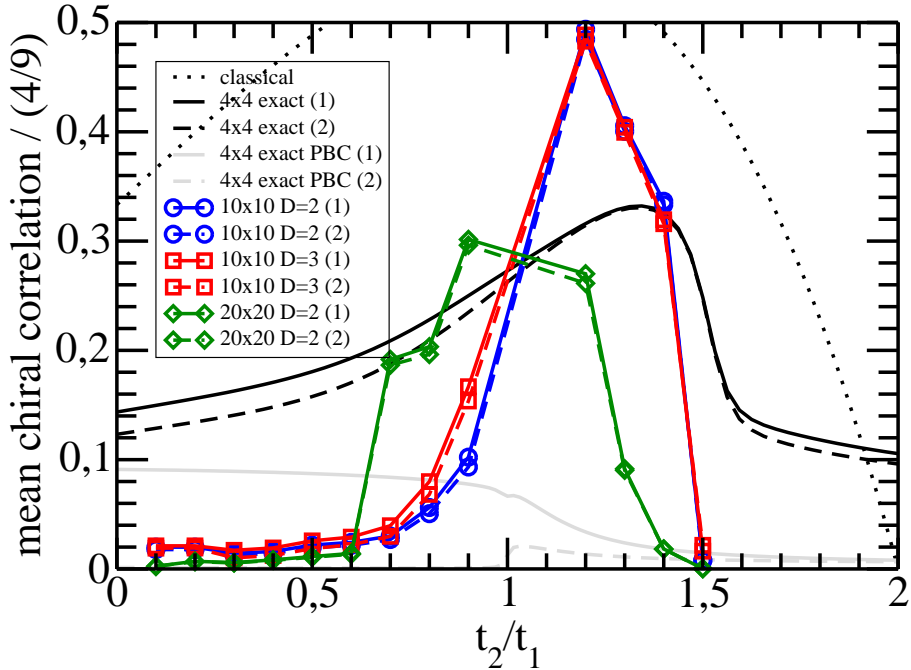}
	\caption{Ground-state vector-chirality correlations $\psi_-/\hat{\psi}_-$ from~\eref{eq:psiminus}, given in terms of the theoretical maximum $\hat{\psi}_-=4/9$. Plotted here are the correlations with the central plaquette pair, averaged over the simulation volume. Dotted line: classical result, which vanishes only in the N\'eel-ordered phase at $t_2/t_1\ge2$. Solid lines (1): plaquette pairs share a $t_1$ link; dashed lines (2): plaquette pairs share a $t_2$ link. Black (no symbols): exact diagonalization (Arnoldi) on a 4x4 lattice. Blue (circles), red (squares): PEPS calculations on a 10x10 lattice with $D=2$ and $D=3$, respectively, demonstrating the adequacy of $D=2$ calculations. Green (diamonds): PEPS 20x20 with $D=2$. Points around $t_2/t_1=1$, where the system is maximally frustrated, have been excluded from the PEPS results due to poor convergence. Grey lines: exact results with periodic boundary conditions, for comparison.}
	\label{fig:totalchirality}
\end{figure}

At the classical level, varying $\vartheta$ from zero continuously deforms the three-sublattice structure of the ground state of the isotropic triangular antiferromagnet, shifting the peaks in the structure factor away from the corners of the first Brillouin zone. For both zero and nonzero $\vartheta$, the classical ground state has a finite vector chirality, defined on a plaquette with counter-clockwise labeled corners $(i,j,k)$ as \cite{Kawamura2002}
\begin{equation}
	\label{eq:kappa3}
	\kappa_{\bigtriangleup} = \frac{2}{3\sqrt{3}}[\vect{S}_i\times\vect{S}_j + 
	\vect{S}_j\times\vect{S}_k + \vect{S}_k\times\vect{S}_i]_z,
\end{equation}
and long-range chirality correlations, defined on plaquette pairs as \cite{Richter1991}
\begin{equation}
	\label{eq:psiminus}
	\psi_- = \langle (\kappa_{\bigtriangleup}-\kappa_{\bigtriangledown})
	(\kappa_{\bigtriangleup'}-\kappa_{\bigtriangledown'})
	\rangle,
\end{equation}
where the plaquette pairs $(\bigtriangleup,\bigtriangledown)$ and $(\bigtriangleup',\bigtriangledown')$ each share two spins. We find that even in the quantum model it is numerically irrelevant which edge is being shared (along the \dirA\ or a \dirB\  direction), as shown in \fref{fig:totalchirality}.

In the quantum limit $S=\frac12$ we neglect the long-range dipolar tail of interactions, restricting our attention to the case of nearest-neighbor intra-chain ($t_1$) and inter-chain ($t_2$) couplings. According to the previous discussion on the dependence of inter-chain couplings upon the angle $\vartheta$, one can experimentally span the range of parameter ratios $0 < |t_2/t_1| < 5/4$; the sign of $t_2/t_1$ is of no relevance since it can be reversed by a $\pi$ rotation of every other \dirA-row of spins [with a corresponding modification to~\eref{eq:kappa3}]. We have done calculations up to $t_2/t_1=3/2$, which may be experimentally accessible by minor deformations of the Wigner crystal. Numerical treatment of the two-dimensional model with exact diagonalization is restricted to very small lattices, while other approaches such as quantum Monte Carlo are hindered by the sign problem due to frustration. We hence resort to a variational calculation based on projected-entangled pair states (PEPS) \cite{Verstraete2004,Murg2007}, which allows us to look at much larger lattices than with exact diagonalization. The calculations are performed on rhombic 10x10 and 20x20 lattices with open boundary conditions; such boundary conditions, while being essentially inherent to the PEPS method, are particularly welcome in a system developing incommensurate spiraling order due to frustration: in \fref{fig:totalchirality}, exact results on a 4x4 lattice with \emph{periodic} boundary conditions are shown to exhibit much reduced chiral order. Moreover, they mimic the natural boundary conditions realized experimentally in a finite Coulomb crystal. Due to the significant scaling of the computational effort with the bond dimension $D$ of the variational PEPS basis, we limit ourselves to the smallest bond dimensions ($D=2,3$) going beyond the mean-field limit $D=1$. Despite this aspect we still capture dramatic quantum features due to the interplay between frustration and quantum fluctuations.

\begin{figure}
	\centering
	\includegraphics[width=8cm]{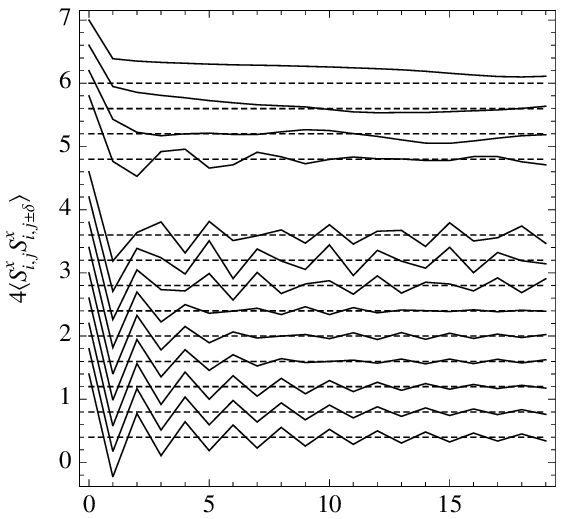}\\
	\includegraphics[width=8cm]{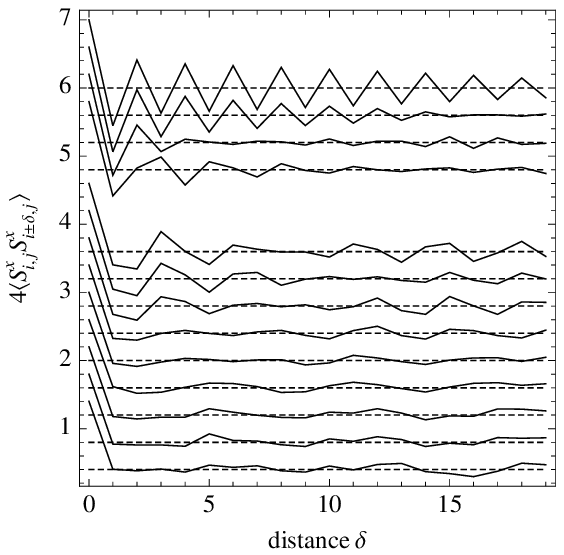}
	\caption{Ground-state spin--spin correlation functions $4\langle S_{i,j}^x S_{i+\delta_i,j+\delta_j}^x\rangle$ along the \dirA-direction (top, along $t_1$ bonds) and along the \dirB\ direction (bottom, along $t_2$ bonds), averaged over the simulation volume as in~\eref{eq:avgspinspin}. Values of $t_2/t_1$ are \{0.1, 0.2, 0.3, 0.4, 0.5, 0.6, 0.7, 0.8, 0.9, 1.2, 1.3, 1.4, 1.5\} from bottom to top, with the dashed lines indicating the shifted reference lines of zero correlations. Points around $t_2/t_1=1$, where the system is maximally frustrated, have been excluded due to poor convergence of the PEPS calculations. Simulation size was $20\times20$ with bond dimension $D=2$.}
	\label{fig:3structurefactor}
\end{figure}

\begin{figure}
	\centering
	\includegraphics[width=12cm]{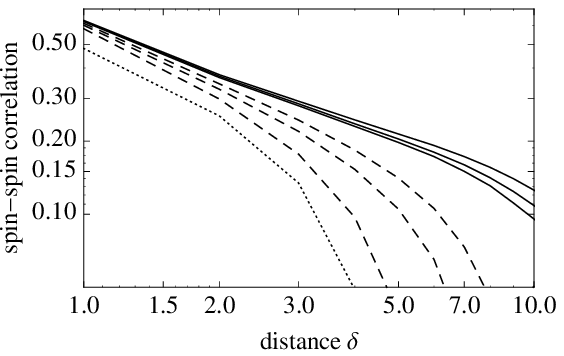}
	\caption{Top six lines: comparison of the lowest six curves of the upper panel of \fref{fig:3structurefactor}: \dirA\ structure factors $4(-1)^{\delta}\langle S_{i,j}^xS_{i,j\pm\delta}\rangle$ for  $t_2/t_1=$ \{0.1, 0.2, 0.3, 0.4, 0.5, 0.6\} from top to bottom. Fits to the short-range part ($2\le\delta\le5$) reveal a transition from polynomial (solid lines) to exponential (dashed lines) decay around $t_2/t_1\approx0.4$. Dotted line: \dirB\ structure factor $4(-1)^{\delta}\langle S_{i,j}^xS_{i\pm\delta,j}\rangle$ for $t_2/t_1=1.3$, showing exponential decay.}
	\label{fig:3structurefactorlog}
\end{figure}

In \fref{fig:totalchirality} we report the evolution of chiral order upon increasing the spatial anisotropy in the couplings. We observe strong indications for the breakdown of chiral order around $t_2/t_1 \approx 0.6$ and $t_2/t_1 \approx 1.3$ via quantum phase transitions. Indeed, at the classical level long-range chiral order is expected for all $t_2/t_1<2$, as it is associated with the a spiral state with a well defined helicity of the spirals; chiral order is expected to vanish only for $t_2/t_1\ge2$ (square lattice XY antiferromagnet). Quantum mechanically, the limits $t_2=0$ and $t_2/t_1\to\infty$ are well known to give a spin liquid with algebraically decaying correlations, and long-range N\'eel order \cite{Dyson1978}, respectively. The way these two limits are attained is nonetheless highly non-trivial, as clearly shown by the evolution of spin--spin correlations upon changing $t_2/t_1$. In the following we will focus on the spin--spin correlations along the \dirA\ direction (of the $t_1$ couplings) and along the \dirB\ direction (of the $t_2$ couplings). In particular, we indicate with $(i,j)$ the position of a point in the $(\dirA,\dirB)$ reference system, and for any inter-site separation $\delta$ we average over all pairs of sites at a distance of $\delta$, along \dirA\ and \dirB. For instance, for correlations along \dirA, we define in analogy to~\eref{eq:chiralorder_a}
\begin{equation}
	\label{eq:avgspinspin}
	\langle S_{i,j} S_{i+\delta,j} \rangle = 
	\frac{1}{N_{\delta}} \sum_{(i,j)_{\delta}}
	\langle S_{i,j} S_{i+\delta,j} \rangle,
\end{equation}
where $N_{\delta}$ is the number of sites $(i,j)_{\delta}$ in the sum for which $(i+\delta,j)$ is also within the simulation cell. \Fref{fig:3structurefactor} shows the spin--spin correlations for different values of $t_2/t_1$. The correlations along the \dirB\ axis build up very slowly upon increasing $t_2/t_1$, and they become significant only for $t_2/t_1\approx 0.7$, consistently with the appearance of chiral order as observed in \fref{fig:totalchirality}. Hence in the interval $0 < t_2/t_1 \lesssim 0.6$ the interplay between quantum fluctuations and frustration leads to an effective decoupling between  the \dirA\ chains. What is then the state along each \dirA-chain? Its evolution with increasing $t_2/t_1$ is absolutely non-trivial and completely dominated by quantum effects. In fact, classically a staggered ordering appears along each chain for $t_2=0$ and  it turns continuously  into (generally incommensurate) spiraling order for any finite $t_2$. The spiral order reproduces the three-sublattice structure of an isotropic triangular lattice for $t_2=t_1$, and it realizes N\'eel order on the $t_2$-lattice when $t_2/t_1\ge2$. This evolution is sketched in \fref{fig:PhD}(a).   Quantum mechanically, on the other hand,  for $t_2/t_1 \ll 1$ the dominant correlations along each chain remain antiferromagnetic, and  the decay of correlations becomes \emph{stronger} for larger $t_2$: as shown in \fref{fig:3structurefactorlog}, the decay is initially of the type $1/r^{K/2}$ with $K=1$ at $t_2=0$. $K$ is seen to grow upon increasing $t_2$: the two-dimensional system hosts a one-dimensional Luttinger-liquid state on each of the \dirA-chains, with a non-universal $K$ exponent. Even more strikingly,  the algebraically decaying staggered correlations turn to exponential decay for $0.4\lesssim t_2/t_1 \lesssim  0.6$, where the state on each chain evolves from a gapless to a gapped spin liquid. Only upon increasing  $t_2/t_1$ further we observe that long-range correlations build up again, this time with a spiraling structure.
 
When increasing $t_2/t_1$ above 1, a similarly non-trivial evolution of correlations takes place. We observe that the loss of chiral order, occurring at  $t_2/t_1\approx 1.3$, is accompanied by a loss in correlations along the \dirB\ direction, rendering the \dirA-chains again effectively decoupled. The state along each \dirA-chain appears to be again a gapped spin liquid, with exponentially decaying correlations. It is only for $t_2/t_1 \approx 1.5$ that the system exhibits long-range correlations in both the \dirA\ and \dirB\ directions, the latter of which already exhibits a full N\'eel structure. \Fref{fig:PhD}(b) summarizes our quantum phase diagram based on PEPS calculations, which exhibit three spin-liquid phases: a gapless one and two gapped ones. The inclusion of long-range dipolar interactions, which physically arise in the ion system, is at present technically challenging; however, we expect the long tail of interactions to give only small corrections to phases which develop a finite gap in the spectrum or finite long-range order in the model limited to short-range couplings, as is the case for most phases in \fref{fig:PhD}(b).

The study of quantum antiferromagnetism on the anisotropic triangular   lattice has recently attracted a deep interest in the case of   Heisenberg interactions, in connection with the physics  of quasi-2D antiferromagnets such as Cs$_2$CuCl$_4$ \cite{Coldea2001}. In particular, a similar phase diagram to the one  presented in \fref{fig:PhD}(b) has been obtained in \cite{Yunoki2006} by variational calculations.   It is extremely intriguing to observe that one  of the most striking features of the phase diagram  predicted for the Heisenberg case \cite{Yunoki2006}, namely the emergence of  intermediate gapped spin-liquid phases in the system,  is also present in the case of XY interactions.   Hence the evolution from   a 1D gapless spin liquid to a spiraling ordered state,  and from spiral to N\'eel order, apparently acquires  a ``universal'' discontinuous structure: instead of a   continuous deformation of correlations  in the ground state, exhibited by the classical system,  the quantum system first shows a complete loss of   (quasi-)long-range correlations in favor of a short-range  spin-liquid state, and then a revival of correlations  at a different wavevector.

\subsection{Preparation of the half-filled system}

The results for the spin models in this section assume a ground state with zero total magnetization, which corresponds to a phonon model with population $N/2$ assuming $N$ active degrees of freedom. Such a state may be more easily prepared in the absence of anharmonicities, as a BEC of phonons described in \sref{sec:BEC}. The optical lattice constituting the microtraps is subsequently switched on adiabatically, which transforms the BEC into the half-filled ground state of the Bose--Hubbard Hamiltionian~\eref{eq:BHham} and, for very large anharmonicities, into that of the $S=\frac12$ spin model.

The phonon BEC at the starting point of this adiabatic passage cannot be prepared by cooling a thermal population of phonons, since their number is not conserved during the cooling process. Instead, we propose that the harmonic trap holding the ion crystal is opened to the point where just the lowest-frequency vibrational mode becomes soft. In a linear chain, for example, we open the radial confinement to the point where the chain acquires a small zig-zag amplitude \cite{Birkl1992,Schiffer1993}. The ion crystal is then cooled to its vibrational ground state in the new symmetry-broken configuration, after which the trap is rapidly closed to its original state. Depending on the amplitude of this symmetry breaking, the final state of the system is a coherent state of phonons in the lowest-frequency mode. In a sufficiently large system, adjusting the mean number of phonons to the half-filling point results in a coherent state which is not very far from a Fock state of exactly half filling. A systematic discussion of this procedure is forthcoming \cite{SchmiedLeibfried}.

\section{Discussion and Conclusions}

In this paper we have presented a variety of quantum many-body Hamiltonians realized by the anharmonic vibrations of ions trapped in the wells of a deep optical lattice. While the dominant Coulomb interaction and the overall trapping potential determine the geometry of the ionic Wigner crystal, at the same time the optical lattice allows to engineer the phononic states for the ions, and it endows the system with a broad range of tunability. On the one hand one can control the non-linearity of phonons (their effective on-site repulsion/attraction) by modifying the optical lattice or superlattice potential. In this way one can explore a regime of essentially harmonic phonons, which can mediate anisotropic spin--spin interactions between internal degrees of freedom of the ions \cite{Porras2004}; a regime of weakly interacting phonons, which, in the case of a two-dimensional triangular-lattice Wigner crystal, condense in a non-trivial finite-momentum state containing vortex-antivortex pairs [\fref{fig:PhD}(a)]; and a regime of essentially hard-core phonons, which allows a mapping of the phononic Hamiltonian onto a frustrated $S=\frac12$ XY model with spatially anisotropic couplings, tuned by changing the orientation of the optical lattice with respect to the ionic crystal. In this latter case we show that an extremely rich physical picture can be obtained in the case of a half-filled system of phonons: in particular, one can realize a gapless spin-liquid phase with long-range chiral order in zig-zag ion ladders, and a variety of two-dimensional spin-liquid phases, either gapped or gapless, in the case of a triangular Wigner crystal. Hence this setup offers a platform for the exploration of highly non-trivial quantum phases and quantum critical phenomena in tunable frustrated quantum magnets.

As an outlook, it is important to realize that an optical lattice potential is just a simple instance of a bigger family of microtraps, in which each ion is trapped individually in a tunable local potential. In the case of an optical lattice all individual trapping potentials are typically identical, so that only global tunability is possible. On the other hand it is possible to fabricate independent microtraps  for the ions \cite{Cirac2000,Kielpinski2002,Stick2006,Pearson2006,Seidelin2006,Chiaverini2007,Schulz2007}, which guarantees at the same time individual tunability of the phononic states of each ion independently, and the possibility of realizing ion-lattice geometries which go beyond the natural Wigner crystals, and even beyond the paradigm of translationally invariant arrays. Work is in progress \cite{SchmiedLeibfried} to identify the quantum many-body Hamiltonians that can be realized in the context of microfabricated traps.

\ack

This work was supported by the European Union through the SCALA integrated project.

\section*{References}

\bibliography{MPQ}

\end{document}